# Nonlinear Schrödinger equation containing the time derivative of the probability density: A numerical study

**Ji Luo**

*Department of Photoelectric Technology,*
*Chengdu University of Information Technology, Chengdu 610225, Sichuan, China*
*Surface Physics Laboratory, Department of Physics, Fudan University,*
*Shanghai 200433, China*

## Abstract

The simplest nonlinear Schrödinger equation that contains the time derivative of the probability density is investigated. This equation has the same stationary solutions as its linear counterpart, and these solutions are the eigenstates of the corresponding linear Hamiltonian. The equation leads to the usual continuity equation and thus maintains the unitarity of the wave function. For the non-stationary solutions, numerical calculations are carried out for the one-dimensional infinite square-well potential and for several time-dependent potentials that tend to the former as time increases. Results show that for various initial states, the wave function always evolves into some eigenstate of the corresponding linear Hamiltonian of the one-dimensional infinite square-well potential. For a small time-dependent perturbation potential, solutions present the process similar to the spontaneous transition between stationary states. For a periodical potential with an appropriate frequency, solutions present the process similar to the stimulated transition. This nonlinear Schrödinger equation thus presents the state evolution similar to the wave-function reduction.

## 1. Introduction

In standard quantum mechanics, the wave-function reduction is introduced to interpret the quantum measurement [1,2]. As a quantity is measured for a quantum system, the wave function collapses into one of the eigenstates of the corresponding operator. This reduction cannot be described by the standard linear Schrödinger equation (LSE), because in general the final state should be the superposition of those eigenstates. Why a single eigenstate is finally realized is the central problem of the quantum measurement [3]. Enlightening researches have demonstrated that the nonlinear Schrödinger equation (NLSE) may play an important role in describing the reduction, especially for the combination of the microscopic system, the macroscopic apparatus, and the environment [4-8]. In the original meaning of the wave-function reduction, however, details are still unclear, and the reduction is generally assumed to be an instantaneous, indeterministic, and irreversible process [1,2,9].

The NLSE containing the probability density [10-15] has been being widely investigated for its applications in nonlinear optics [10,16-18], in plasma physics [19], and to the Bose-Einstein condensates [20,21,22]. The most important feature of this equation is the solutions in the form of solitons. In quantum mechanics, the NLSE containing the probability density varies the eigen energies of the Hamiltonian corresponding to the LSE [23]. The NLSE containing the derivative of the probability density with respect to the space coordinates also has applications in plasma physics and nonlinear optics [24]. The NLSE containing the derivative of the probability density with respect to the time coordinate, however, is less popular in the literature and mainly appears in the field of nonlinear optics [25]. In this paper, it is demonstrated that to exactly describe the wave-function variation of a quantum particle, the time derivative of the probability density should be involved in the equation of dynamics. The simplest NLSE that contains this time derivative is investigated. This NLSE has the same stationary solutions as its linear counterpart, and these stationary solutions are the eigenstates of the corresponding linear Hamiltonian. Besides, this NLSE leads to the usual continuity equation and thus maintains the unitarity of the wave function. To obtain non-stationary solutions, numerical calculations are carried out for the one-dimensional infinite square-well potential (1D ISWP) and for several time-dependent potentials that tend to the 1D ISWP as time increases. Results show that for various initial states, the wave function always evolves into some stationary state that is an eigenstate of the linear 1D ISWP Hamiltonian. This NLSE thus presents the state evolution similar to the wave-function reduction. For a small time-dependent perturbation potential, the NLSE presents the process similar to the spontaneous transition between stationary states. For a periodical potential, the NLSE presents the process similar to the stimulated transition, if the frequency of the potential equals the energy difference of two stationary states.

The remaining part of the article is arranged as follows: In section 2, inexactness of the standard LSE is discussed. In section 3, the simplest NLSE containing the time derivative of the density is investigated. In Section 4, numerical solutions of this NLSE for the 1D ISWP are presented. In Section 5, the time-dependent potentials are considered. In Section 6, we conclude the present work.

## 2. Inexactness of the standard linear Schrödinger equation

Suppose a particle with mass $m_0$ and charge $q$ moves in the external electromagnetic field $\vec{E}$, $\vec{B}$. The field is also described by the scalar potential $\varphi(\vec{r},t)$ and the vector potential $\vec{A}(\vec{r},t)$. In quantum mechanics, the Hamiltonian of the particle,

$$\hat{H} = \frac{1}{2m_0}(i\hbar\nabla + q\vec{A})^2 + q\varphi\,, \tag{1}$$

is obtained according to its non-relativistic classical counterpart [26,27]

$$H(\vec{r},\vec{P}) = \frac{1}{2m_0}(\vec{P} - q\vec{A})^2 + q\varphi\,. \tag{2}$$

The way is to substitute the canonical momentum $\vec{P} = m_0\vec{v} + q\vec{A}$ in Eq. (2) with the operator $\hat{\vec{P}} = -i\hbar\nabla$, where $\vec{v}$ is the velocity. Further backward, the Hamiltonian of the classical particle is derived from the Lagrangian $L(\vec{r},\vec{v}) = m_0 v^2/2 - q(\varphi - \vec{v}\cdot\vec{A})$. Finally, the Lagrangian has this form because the Lagrange equation $(d/dt)(\partial L/\partial v_w) - \partial L/\partial w = 0$ with $w = x, y, z$ for the classical particle is equivalent to the Newton equation

$$m_0\frac{d\vec{v}}{dt} = q(\vec{E} + \vec{v}\times\vec{B})\,. \tag{3}$$

Hence procedurally, the Hamiltonian operator $\hat{H}$ originates from the Lorentz force that governs the motion of the classical particle.

The charged classical particle radiates electromagnetic wave if it moves in acceleration. However, Eq. (3) does not include the damping effect due to this radiation. Consequently, Eq. (2) is an approximation and so is its quantum mechanics counterpart, Eq. (1). This means that the standard LSE

$$i\hbar\frac{\partial\psi}{\partial t}=\hat{H}\psi\,, \tag{4}$$

where $\psi(\vec{r},t)$ is the wave function, does not exactly describe the state variation of the quantum particle because effects of the radiation are not included. While Eq. (3) describes the motion of the classical particle very well, the radiation may affect the state variation of the quantum particle quite differently.

The charged quantum particle generates electromagnetic field of its own. If the particle stays in a bound state $\psi$, this field at distant positions can be described by the following retarded potentials [27]:

$$\varphi_q(\vec{r},t)=\frac{q}{4\pi\varepsilon_0}\int_\infty\frac{\rho(\vec{r}\,',t-|\vec{r}-\vec{r}\,'|/c)}{|\vec{r}-\vec{r}\,'|}d\vec{r}\,'\,, \tag{5}$$

$$\vec{A}_q(\vec{r},t)=\frac{\mu_0 q}{4\pi}\int_\infty\frac{\vec{J}(\vec{r}\,',t-|\vec{r}-\vec{r}\,'|/c)}{|\vec{r}-\vec{r}\,'|}d\vec{r}\,'\,, \tag{6}$$

where $\rho(\vec{r},t)=\psi(\vec{r},t)\psi^*(\vec{r},t)$ is the probability density and $\vec{J}(\vec{r},t)=-(i\hbar/2m)(\psi^*\nabla\psi-\psi\nabla\psi^*)-(q/m)\vec{A}\,\psi\psi^*$ is the current density. The asterisk denotes complex conjugate throughout this paper. The remote field is given by $\vec{E}_q=-\nabla\varphi_q-\partial\vec{A}_q/\partial t$ and $\vec{B}_q=\nabla\times\vec{A}_q$.

If the particle stays in a stationary state $\psi(\vec{r},t)=\phi(\vec{r})\exp[i\theta(t)]$, where $\theta(t)$ is a real function, both $\rho$ and $\vec{J}$ are independent of time. Hence $\vec{E}_q$ and $\vec{B}_q$ are also stationary. This corresponds to the case without radiation. For a general wave function $\psi(\vec{r},t)$, both $\rho$ and $\vec{J}$ vary with time, and the particle radiates electromagnetic wave described by Eqs. (5) and (6) at distant positions. It is thus a reasonable argument that derivatives of $\rho$ and $\vec{J}$ with respect to time should appear in the more exact equation of dynamics, if effects of the radiation on the wave-function evolution are to be included. The NLSE containing the time derivative of the probability density thus becomes a necessary and interesting topic.

## 2. The simplest nonlinear Schrödinger equation containing the time derivative of the probability density

The simplest NLSE that contains the time derivative of the probability density seems to be

$$i\frac{\partial\psi}{\partial t}=-\alpha\nabla^2\psi+V_0(\vec{r})\psi+\beta\frac{\partial(\psi\psi^*)}{\partial t}\psi\,, \tag{7}$$

where $V_0(\vec{r})$ is a time-independent potential, $\alpha>0$ and $\beta$ are real parameters. Values of $\alpha$ and $\beta$ depend on the system of units. Roughly speaking, $\alpha$ is related to the mass of the particle and $\beta$ represents the intensity of nonlinear effects. For instance, $\alpha=0.5$ corresponds to the electron in atomic units. For $\beta=0$, we recover the standard LSE. In Eq. (7),

$$\hat{H}_0 = -\alpha\nabla^2 + V_0(\vec{r}) \tag{8}$$

is the usual linear Hamiltonian of the particle. Equation (7) leads to the same continuity equation as the LSE of $\hat{H}_0$, that is,

$$\frac{\partial\rho}{\partial t} = -\nabla\cdot\vec{J}\,, \tag{9}$$

where $\vec{J} = -i\alpha(\psi^*\nabla\psi - \psi\nabla\psi^*)$ is the ordinary current density. Equation (7) thus maintains the unitarity of the wave function. Due to Eq. (9), Eq. (7) can also be expressed as

$$i\frac{\partial\psi}{\partial t} = -\alpha\nabla^2\psi + V_0(\vec{r})\psi + i\alpha\beta(\psi^*\nabla^2\psi - \psi\nabla^2\psi^*)\psi\,. \tag{10}$$

Depending on boundary conditions, $\hat{H}_0$ has real eigenfunctions $\phi_n(\vec{r})$ and corresponding eigenvalues $E_n$ that satisfy

$$\hat{H}_0\phi_n = E_n\phi_n \tag{11}$$

with $n = 1,2,\cdots$. For simplicity we suppose that all $E_n$ are non-degenerate. Hence all $\phi_n(\vec{r})$ are independent and constitute a complete set of orthonormal functions. Equation (7) has the stationary solutions

$$\psi_n(\vec{r},t) = \phi_n(\vec{r})\exp(-iE_n t) \tag{12}$$

with $n = 1,2,\cdots$, since $\psi_n\psi_n^*$ are independent of time and for every $\psi_n$, Eq. (7) reduces to the standard LSE. In fact, except for phase factors $\exp(i\theta_n)$ where $\theta_n$ are real constants, $\psi_n(\vec{r},t)$ with $n = 1,2,\cdots$ are all possible stationary solutions of Eq. (7). Hence Eq. (7) and its linear counterpart have the same stationary solutions.

A general bound state $\psi(\vec{r},t)$ determined by Eq. (7) can be expanded in terms of functions $\phi_n(\vec{r})$ with $n = 1,2,\cdots$, that is,

$$\psi(\vec{r},t) = \sum_{n=1}^{\infty} C_n(t)\phi_n(\vec{r})\,, \tag{13}$$

where $C_n(t)$ with $n = 1,2,\cdots$ are coefficients. One substitutes Eq. (13) into Eq. (7) and obtains $i\sum_{n=1}^{\infty}\phi_n dC_n/dt = \sum_{n=1}^{\infty}\phi_n E_n C_n + \beta\sum_{l=1}^{\infty}\sum_{m=1}^{\infty}\sum_{n=1}^{\infty}\phi_l\phi_m\phi_n C_l d(C_m C_n^*)/dt$. One may then expand the products $\phi_l(\vec{r})\phi_m(\vec{r})\phi_n(\vec{r})$ with $l,m,n = 1,2,\cdots$ in terms of $\phi_k(\vec{r})$ with $k = 1,2,\cdots$, that is, $\phi_l(\vec{r})\phi_m(\vec{r})\phi_n(\vec{r}) = \sum_{k=1}^{\infty} D_{k,l,m,n}\phi_k(\vec{r})$, where

$$D_{k,l,m,n} = \int_{\infty}\phi_k(\vec{r})\phi_l(\vec{r})\phi_m(\vec{r})\phi_n(\vec{r})d\vec{r} \tag{14}$$

with $k,l,m,n = 1,2,\cdots$ are real coefficients. Finally one obtains

$$i\frac{dC_k}{dt} = E_k C_k + \beta\sum_{l=1}^{\infty}\sum_{m=1}^{\infty}\sum_{n=1}^{\infty} D_{k,l,m,n} C_l \frac{d(C_m C_n^*)}{dt}\,, \tag{15}$$

where $k = 1,2,\cdots$. Obviously $D_{k,l,m,n}$ is independent of the order of the subscripts. For a pair of numbers $j,k \in \{1,2,\cdots\}$, one multiplies the complex conjugate of Eq. (15) for $k$ by $C_j$ and multiplies the Eq. (15) for $j$ by $C_k^*$. A subtraction then leads to

$$\sum_{m=1}^{\infty}\sum_{n=1}^{\infty}\left[i\delta_{j,m}\delta_{k,n} + \beta\sum_{l=1}^{\infty}\left(D_{k,l,m,n}C_j C_l^* - D_{j,l,m,n}C_l C_k^*\right)\right]\frac{d(C_m C_n^*)}{dt} = (E_j - E_k)C_j C_k^*\,, \tag{16}$$

where $j,k=1,2,\cdots$, and $\delta_{j,m}$, $\delta_{k,n}$ are Kronecker functions. This is a set of differential equations of the products $C_m C_n^*$ with $m,n=1,2,\cdots$ and is suitable for numerical treatment. Equation (16) and the initial values $C_m(0)C_n^*(0)$ completely determine functions $C_m(t)C_n^*(t)$ with $m,n=1,2,\cdots$. For the normalized initial wave function, one has

$$\sum_{n=1}^{\infty} C_n(t)C_n^*(t)=1 \tag{17}$$

for all $t\geq 0$. Besides, $(C_m C_n^*)(C_m C_n^*)^*=(C_m C_m^*)(C_n C_n^*)\leq 1$. Hence $|C_m C_n^*|\leq 1$ for $m,n=1,2,\cdots$.

## 4. Numerical solutions for the one-dimensional infinite square-well potential

For the 1D ISWP where $V_0(x)=0$ for $0<x<1$ and $V_0(x)=+\infty$ for other $x$, Eq. (7) becomes

$$i\frac{\partial\psi}{\partial t}=-\alpha\frac{\partial^2\psi}{\partial x^2}+\beta\frac{\partial(\psi\psi^*)}{\partial t}\psi , \tag{18}$$

where $0\leq x\leq 1$ and $\psi(0,t)=\psi(1,t)=0$. With zero boundary-conditions, the linear Hamiltonian $\hat{H}_0=-\alpha\partial^2/\partial x^2$ has orthonormal eigenfunctions

$$\phi_n(x)=\sqrt{2}\sin(n\pi x) \tag{19}$$

and corresponding eigenvalues

$$E_n=\alpha(n\pi)^2 \tag{20}$$

with $n=1,2,\cdots$. Coefficients $D_{k,l,m,n}$ for $k,l,m,n=1,2,\cdots$ are calculated to be

$$\begin{aligned} D_{k,l,m,n}=\frac{1}{2}(&\delta_{0,k+l-m-n}+\delta_{0,k-l+m-n}+\delta_{0,k-l-m+n} \\ &-\delta_{0,k+l+m-n}-\delta_{0,k+l-m+n}-\delta_{0,k-l+m+n}-\delta_{0,k-l-m-n}). \end{aligned} \tag{21}$$

By taking the first $N$ terms of the series in Eq. (13) and the first $N$ terms of the expansion of $\phi_l(\vec{r})\phi_m(\vec{r})\phi_n(\vec{r})$, one obtains from Eq. (16) a set of $N\times N$ first-order differential equations of $N\times N$ functions $C_m(t)C_n^*(t)$ with $m,n=1,2,\cdots,N$. These equations are numerically solved for various normalized initial values $C_m(0)C_n^*(0)$ with standard fourth-order Runge-Kutta method combined with high-accuracy solutions of linear systems. The initial values depend on the state at $t=0$ according to

$$\psi(\vec{r},0)=\sum_{n=1}^{N} C_n(0)\phi_n(\vec{r}) . \tag{22}$$

Calculations are carried out with double-precision FORTRAN programs.

Results show that for every initial state, there is always a product $C_k C_k^*$ with $k\in\{1,2,\cdots,N\}$ such that

$$\lim_{t\to+\infty} C_k(t)C_k^*(t)=1 \tag{23}$$

and for all other products $C_m C_n^*$ with $(m,n)\neq(k,k)$

$$\lim_{t\to+\infty} C_m(t)C_n^*(t)=0 . \tag{24}$$

Hence the wave function evolves into a stationary state $C_k(+\infty)\phi_k(x)$, or except for a phase factor, asymptotically

$$\psi(x,t)=\sqrt{2}\sin(k\pi x)\exp[-i\alpha(k\pi)^2 t] \tag{25}$$

for $t\to+\infty$.

Initial states are catalogued according to the values of $C_n(0)C_n^*(0)$ with $n=1,2,\cdots,N$. This amounts to fixing the modulus of every $C_n(0)$. Typical cases include states where $C_n(0)C_n^*(0)$ are identical for several $n$, states where $C_n(0)C_n^*(0)$ decreases with increasing $n$, states where $C_n(0)C_n^*(0)$ are non-zero only for two or three $n$, states where $C_n(0)C_n^*(0)$ are centered on a specific $n$ like a pulse, and states where $C_n(0)C_n^*(0)$ are randomly chosen. Values of other $C_m(0)C_n^*(0)$ with $m\neq n$ are obtained by randomly choosing the angles of $C_n(0)$ with $n=1,2,\cdots,N$. In most calculations, the time step $h=0.0001$ is used and the cut-off number $N$ is between 10 and 20. In some testing calculations, smaller $h$ and greater $N$ are attempted. In general $h=0.0001$ gives reliable results. Usually $N$ is chosen such that the last five $C_n(0)C_n^*(0)$ with $n=N-4,\cdots,N$ are zero. In other testing calculations, we choose the initial state to be an eigenstate $C_j(0)\phi_j(x)$ with $j\in\{1,2,\cdots,N\}$. In this case, all $C_mC_n^*$ with $m,n=1,2,\cdots,N$ do not vary with time.

Typical results are presented in figure 1. Usually the product $C_kC_k^*$ of the final state may eventually satisfy $|C_kC_k^*-1|<10^{-10}$ and products $C_nC_n^*$ for $n\neq k$ may satisfy $|C_nC_n^*|<10^{-10}$ as time increases (due to the calculation errors, $C_nC_n^*$ can be minus numbers that are nearly zero). Equation (17) is always satisfied in the calculations. The final states $C_k(+\infty)\phi_k(x)$ differ for different initial states.

Parameters $\alpha$ and $\beta$ affect the time needed to realize the final state. A smaller $\alpha$ and a smaller $|\beta|$ lead to a smaller convergent rate. Most calculations are carried out for $\alpha=0.5$ and $\beta=1$. For a fixed initial state, the sign of $\beta$ affects the final state. For $\beta=0$, all $C_nC_n^*$ with $n=1,2,\cdots,N$ do not vary with time. Calculations lead to the solutions of the LSE, that is, $\psi(x,t)=\sum_{n=1}^{N}C_n(t)\phi_n(x)$ where $C_n(t)=C_n(0)\exp(-E_nt)$ for $n=1,2,\cdots,N$. For $\beta>0$, the final state $C_k(+\infty)\phi_k(x)$ satisfies $k\leq n$ for any $n\in\{1,2,\cdots,N\}$ with $C_n(0)\neq 0$. Hence the wave function evolves into a lower eigenstate. For $\beta<0$, the final state satisfies $k\geq n$ for any $n\in\{1,2,\cdots,N\}$ with $C_n(0)\neq 0$ and the wave function evolves into a higher eigenstate. Calculation results for $\beta<0$ may be incorrect because in this case, states $\phi_n$ with $n>N$ are needed to express the wave function. We note that $|\beta|$ should be small for Eq. (7) to correctly describe the wave-function evolution. For small values of $\beta>0$, the wave function demonstrates the same tendency of evolution as for $\beta=1$, but the time needed to realize the final state is longer and so is the CPU time of calculations. Conclusively, the nonlinear term in Eq. (7) with $\beta>0$ always makes the wave function to evolve into the possible lowest eigenstate of the corresponding linear Hamiltonian.

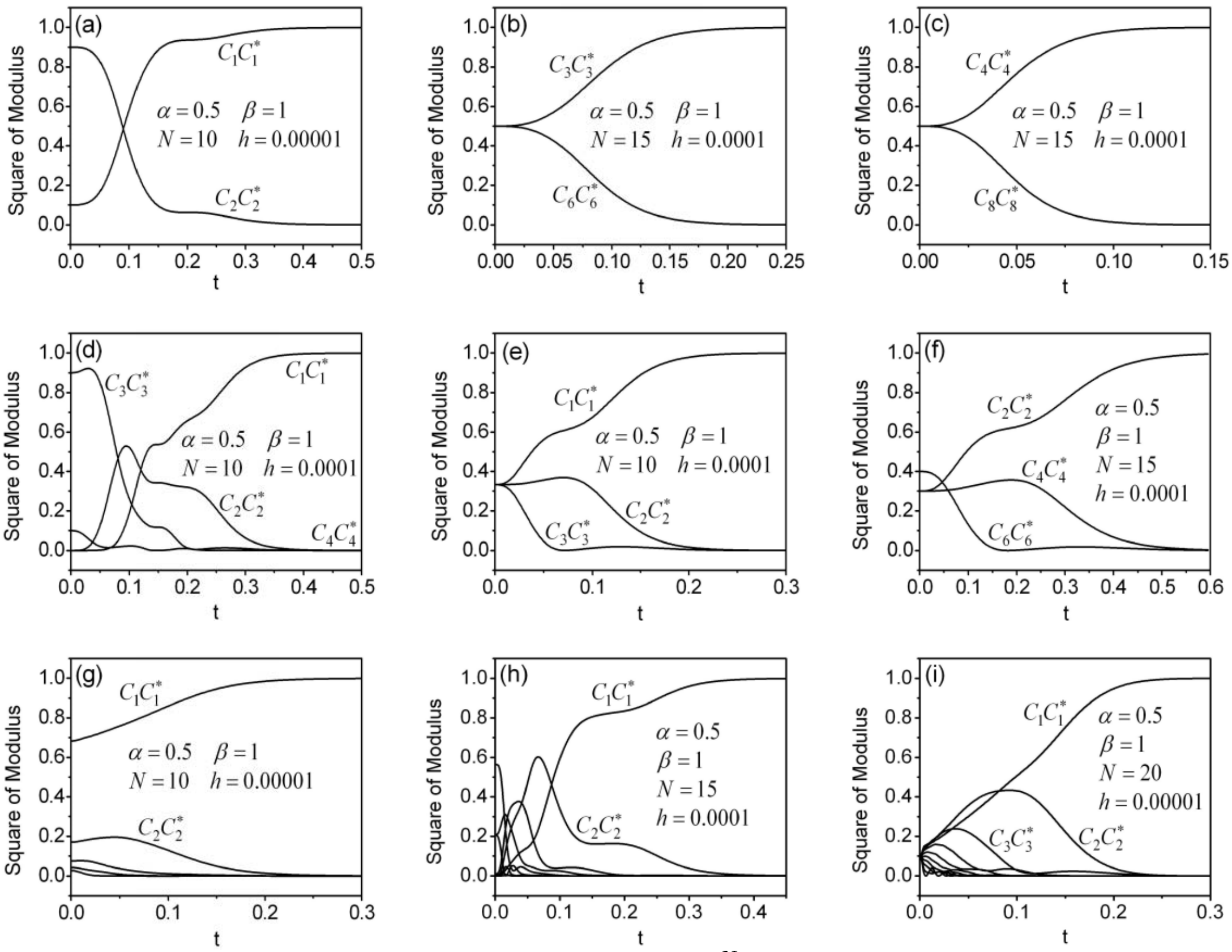

**Figure 1.** Evolution of the wave function $\psi(x,t)=\sum_{n=1}^{N} C_n(t)\phi_n(x)$ represented by $C_nC_n^*$ as functions of time $t$, determined by the NLSE with 1D ISWP. $C_nC_n^*$ with $n=1,2,\cdots,N$ are calculated from Eq. (16) and $\phi_n(x)=\sqrt{2}\sin(n\pi x)$ with $n=1,2,\cdots,N$ are the eigenfunctions of the linear 1D ISWP Hamiltonian. Graphs (a)~(i) correspond to different initial values $C_n(0)C_n^*(0)$ with $n=1,2,\cdots,N$.

## 5. Time-dependent potentials

For a time-dependent potential $V(\vec{r},t)$, the corresponding NLSE is

$$i\frac{\partial\psi}{\partial t}=-\alpha\nabla^2\psi+V(\vec{r},t)\psi+\beta\frac{\partial(\psi\psi^*)}{\partial t}\psi\,. \tag{26}$$

Equation (26) also leads to the continuity equation (9) and maintains the unitarity of the wave function. Suppose the potential depends on time in such a way that

$$\lim_{t\to+\infty}V(\vec{r},t)=V_0(\vec{r})\,. \tag{27}$$

Equation (26) then has asymptotical stationary-solutions expressed in Eq. (12) as $t\to+\infty$. If for any initial state $\psi(\vec{r},0)$, the solution of Eq. (7) evolves into some stationary state, then so should the solution of Eq. (26).

Solutions of Eq. (26) can also be expanded in terms of functions $\phi_n(\vec{r})$ with $n=1,2,\cdots$, as expressed in Eq. (13). Equation (26) corresponds to a time-dependent linear Hamiltonian

$$\hat{H}_t=-\alpha\nabla^2+V(\vec{r},t) \tag{28}$$

that satisfies $\hat{H}_t\phi_n=\hat{H}_0\phi_n+(V-V_0)\phi_n=E_n\phi_n+(V-V_0)\phi_n$. Suppose $V(\vec{r},t)=V_0(\vec{r})$

on the boundaries. One expands the function $V(\vec{r},t)-V_0(\vec{r})$ in terms of $\phi_n(\vec{r})$ with $n=1,2,\cdots$, that is, $V(\vec{r},t)-V_0(\vec{r})=\sum_{n=1}^{\infty}V_n(t)\phi_n(\vec{r})$ where $V_n(t)$ with $n=1,2,\cdots$ are real coefficients that satisfy

$$\lim_{t\to+\infty}V_n(t)=0 \tag{29}$$

according to Eq. (27). Then after substituting Eq. (13) into Eq. (26), one expands the products $\phi_m(\vec{r})\phi_n(\vec{r})$ in terms of $\phi_l(\vec{r})$ with $l=1,2,\cdots$, that is, $\phi_m(\vec{r})\phi_n(\vec{r})=\sum_{l=1}^{\infty}D_{l,m,n}\phi_l(\vec{r})$, where

$$D_{l,m,n}=\int_{\infty}\phi_l(\vec{r})\phi_m(\vec{r})\phi_n(\vec{r})d\vec{r} \tag{30}$$

with $l,m,n=1,2,\cdots$ are coefficients. Like Eq. (16), one obtains

$$\begin{aligned}&\sum_{m=1}^{\infty}\sum_{n=1}^{\infty}\left[i\delta_{j,m}\delta_{k,n}+\beta\sum_{l=1}^{\infty}\left(D_{k,l,m,n}C_jC_l^*-D_{j,l,m,n}C_lC_k^*\right)\right]\frac{d(C_mC_n^*)}{dt}\\&=(E_j-E_k)C_jC_k^*+\sum_{m=1}^{\infty}\sum_{n=1}^{\infty}V_m\left(D_{j,m,n}C_nC_k^*-D_{k,m,n}C_jC_n^*\right)\end{aligned} \tag{31}$$

for $j,k=1,2,\cdots$. Equation (31) can be solved by the same method as Eq. (16). Numerical calculations are carried out for $V_0(\vec{r})$ being the 1D ISWP.

Coefficients $V_n$ completely determine the potential $V(\vec{r},t)$. We take the time-dependent potential according to

$$V_n(t)=\gamma_n t^{\mu}\sin(\omega t+\varphi_0)\exp(-\lambda t), \tag{32}$$

where $n=1,2,\cdots,N$ and $\gamma_n$, $\mu\geq 0$, $\omega\geq 0$, $\varphi_0$, $\lambda\geq 0$ are real constants. For 1D ISWP, coefficients $D_{l,m,n}$ with $l,m,n=1,2,\cdots$ are calculated to be

$$D_{l,m,n}=\frac{\sqrt{2}}{\pi}[-D(l+m+n)+D(l+m-n)+D(l-m+n)-D(l-m-n)], \tag{33}$$

where the function $D(k)=1/k$ for an odd number $k$ and $D(k)=0$ for an even number $k$ with $k=1,2,\cdots$.

First we take $V_n(t)=\gamma_n$, where $\gamma_n$ are randomly chosen. In this case, all $C_mC_n^*$ with $m,n=1,2,\cdots,N$ will eventually not vary with time in the calculations, but in general no $C_nC_n^*$ for $n=1,2,\cdots,N$ will become unit. Hence the wave function evolves into a stationary state that is the superposition of $\phi_n$ with $n=1,2,\cdots,N$. We note that $V_n(t)=\gamma_n$ amounts to a time-independent potential $V(x)\neq V_0(x)$. Hence the final state is a stationary eigenstate of the time-independent linear Hamiltonian corresponding to $V(x)$. This indicates that the main conclusion in Section IV is also correct for more complicated potentials than the 1D ISWP. Typical results are presented in figure 2.

Then we respectively take $V_n(t)=\gamma_n\exp(-\lambda t)$, $V_n(t)=\gamma_n t\exp(-\lambda t)$, and $V_n(t)=\gamma_n\sin(\omega t)\exp(-\lambda t)$. For every initial state, the calculation leads to a stationary final state $C_k(+\infty)\phi_k(x)$ with $k\in\{1,2,\cdots,N\}$. In some cases, a stationary state $C_k(t)\phi_k(x)$ is first realized and the wave function continues to evolve into the ground state $C_1(+\infty)\phi_1(x)$. This process also takes place for a small $\beta>0$, but the time needed is longer. Typical results are presented in figure 3.

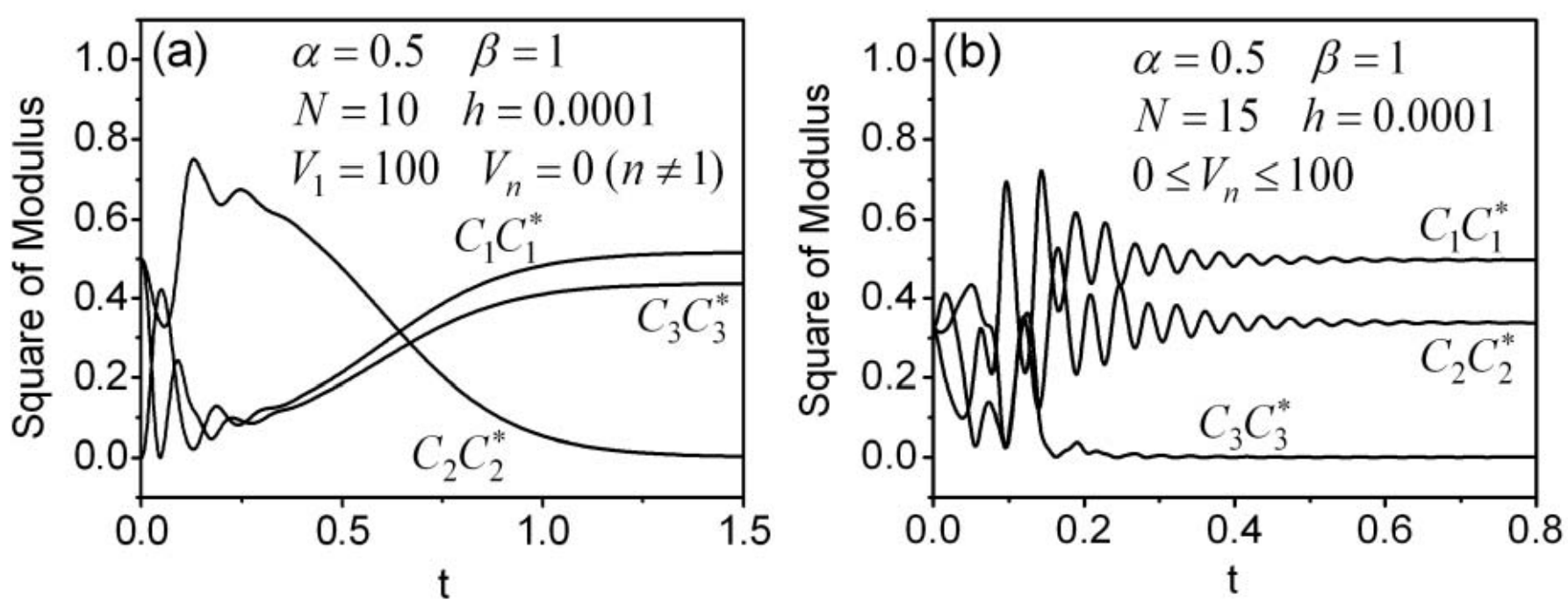

**Figure 2.** Evolution of the wave function $\psi(x,t)=\sum_{n=1}^{N}C_n(t)\phi_n(x)$ represented by $C_nC_n^*$ as functions of time $t$, determined by the NLSE with a time-independent potential $V(x)=\sum_{n=1}^{N}V_n\phi_n(x)$. $C_nC_n^*$ with $n=1,2,\cdots,N$ are calculated from Eq. (31) and $\phi_n(x)=\sqrt{2}\sin(n\pi x)$ with $n=1,2,\cdots,N$ are the eigenfunctions of the linear 1D ISWP Hamiltonian. Graphs (a) and (b) correspond to different initial values $C_n(0)C_n^*(0)$ and different $V_n$ with $n=1,2,\cdots,N$.

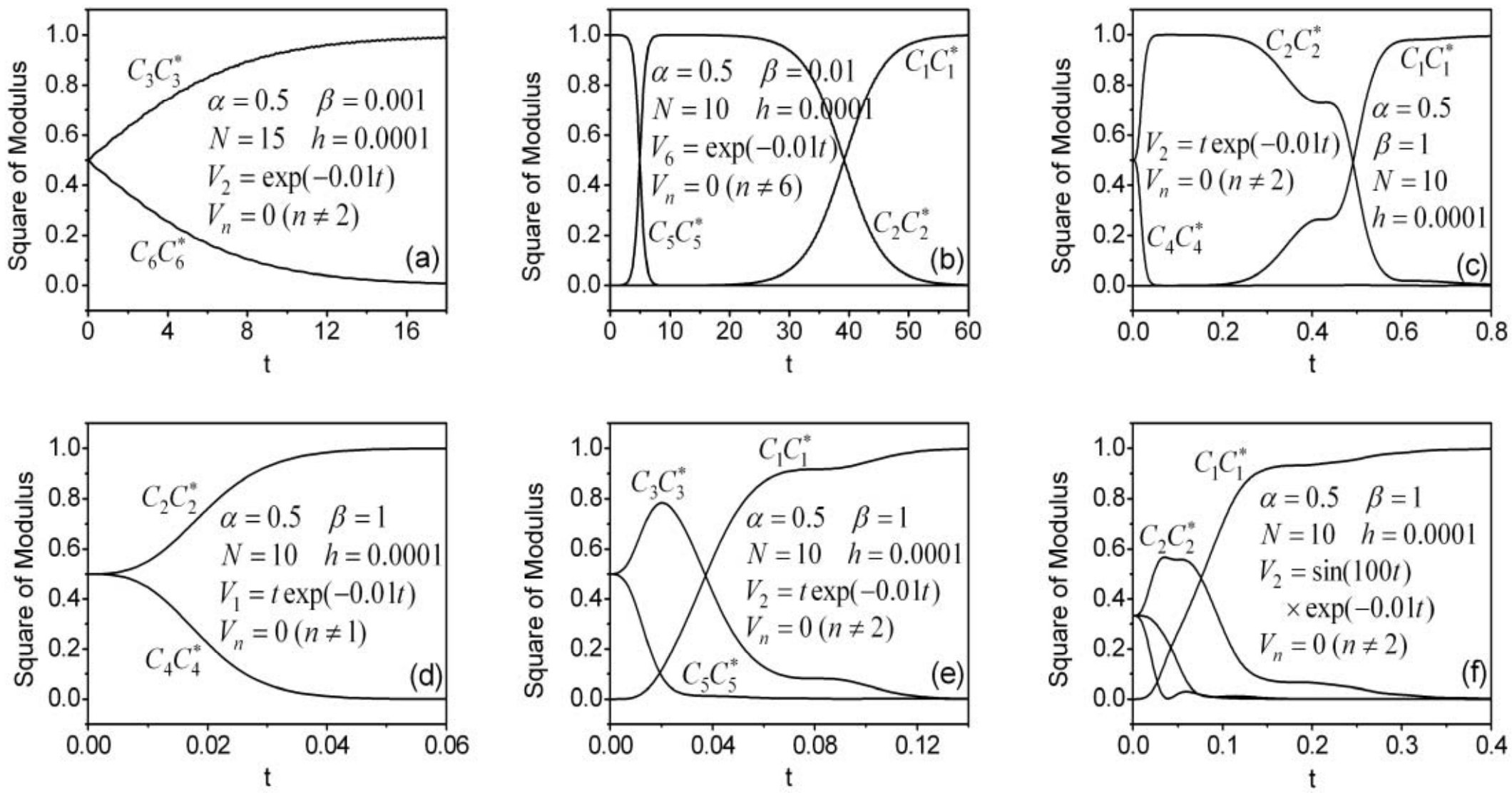

**Figure 3.** Evolution of the wave function $\psi(x,t)=\sum_{n=1}^{N}C_n(t)\phi_n(x)$ represented by $C_nC_n^*$ as functions of time $t$, determined by the NLSE with a time-dependent potential $V(x,t)=\sum_{n=1}^{N}V_n(t)\phi_n(x)$ that tends to 1D ISWP as $t\to+\infty$. $C_nC_n^*$ with $n=1,2,\cdots,N$ are calculated from Eq. (31) and $\phi_n(x)=\sqrt{2}\sin(n\pi x)$ with $n=1,2,\cdots,N$ are the eigenfunctions of the linear 1D ISWP Hamiltonian. Graphs (a)~(f) correspond to different initial values $C_n(0)C_n^*(0)$ with $n=1,2,\cdots,N$ and different $V(x,t)$.

If the initial state is an eigenstate $C_j(0)\phi_j(x)$ with $j \in \{1,2,\cdots,N\}$, calculations present the transition between stationary states. Even a small potential disturbance $V(x,t)-V_0(x)$ may result in the transition from $C_j(0)\phi_j(x)$ to the final state $C_k(+\infty)\phi_k(x)$. For the same initial eigenstate, different potentials lead to different final states. Calculations also demonstrate that sometimes the realized eigenstate $C_k(t)\phi_k(x)$ as an excited state may continue to evolve and the wave function eventually transforms into the ground state $C_1(+\infty)\phi_1(x)$, as long as the perturbation potential persists. Transition may also be realized for a small periodical potential that does not satisfy Eq. (27). Typical results are presented in figure 4.

Finally we take the initial state $C_j(0)\phi_j(x)$ and the coefficients $V_n(t)=\gamma_n \cos(\omega t)$ where $n=1,2,\cdots,N$ and $\omega = | E_k - E_j |$ with $j,k \in \{1,2,\cdots,N\}$. According to ordinary perturbation theory, for the LSE the wave function evolves from the initial state $C_j(0)\phi_j(x)$ in such a way that $C_k(t)C_k^*(t) \to 1$ and $C_n(t)C_n^*(t) \to 0$ for $n \neq k$ as $t \to +\infty$. Our calculations for $\beta = 0$ demonstrate this kind of evolution. For the NLSE, the same results are also obtained from the calculations for small $\beta > 0$. For instance, for $V_n(t)=\cos[(E_3-E_2)t]$, $\beta = 0.0001$, and the initial state $\phi_3$, after $t=3.5$ we have $C_2(t)C_2^*(t) > 0.99$. If the initial state is $\phi_2$, we obtain $C_3(t)C_3^*(t) > 0.99$. For $\beta = 1$, however, the results are $C_1(t)C_1^*(t) \to 1$, although $\omega \neq E_3 - E_1$ and $\omega \neq E_2 - E_1$. Results are presented in figure 5.

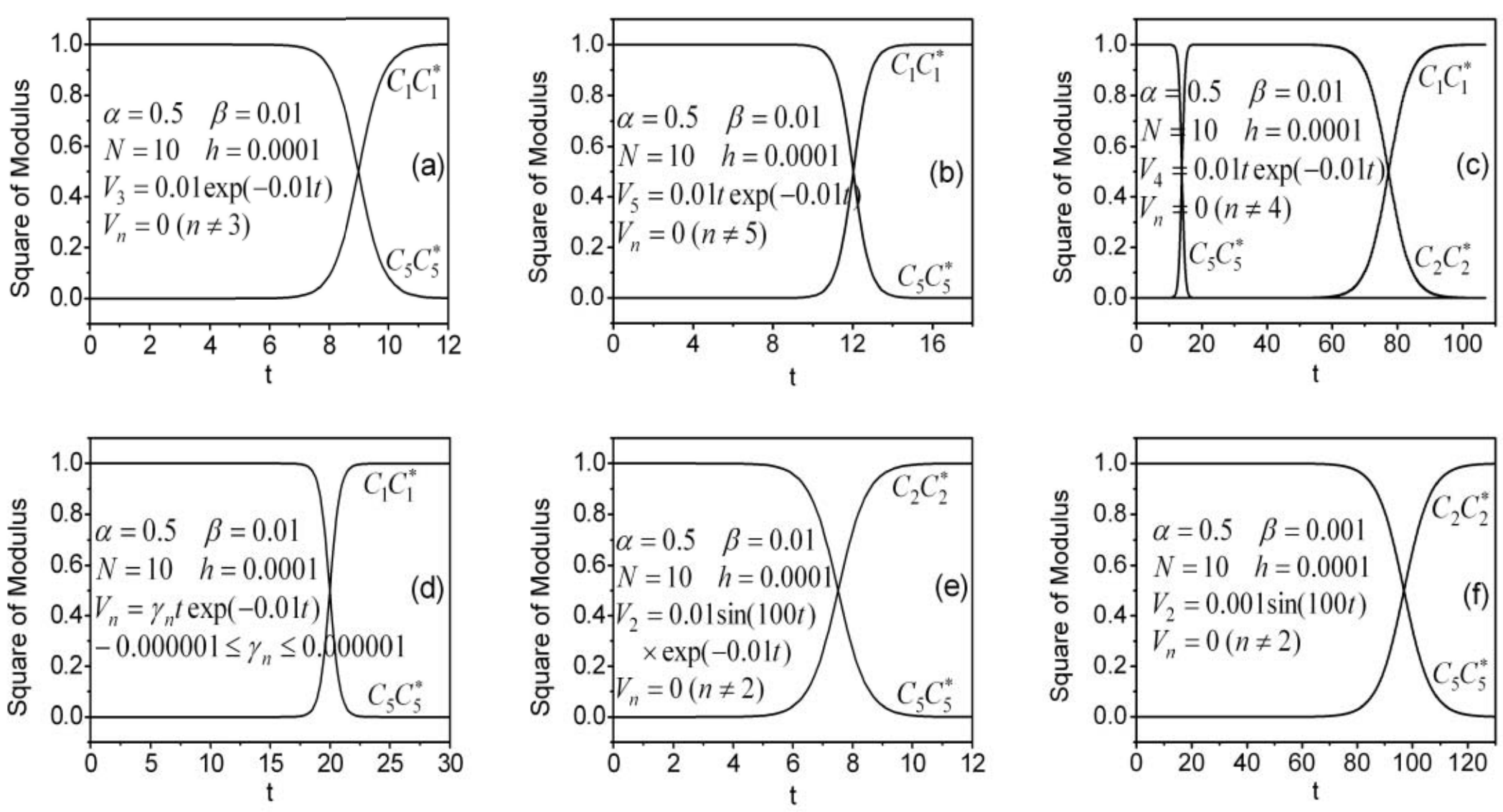

**Figure 4.** Transition between stationary states from the same initial state $\phi_5(x)$, determined by the NLSE with a time-dependent perturbation potential $V(x,t)=\sum_{n=1}^{N} V_n(t)\phi_n(x)$, where $\phi_n(x)=\sqrt{2}\sin(n\pi x)$ with $n=1,2,\cdots,N$ are the eigenfunctions of the linear 1D ISWP Hamiltonian. Graphs (a)~(f) correspond to different potentials.

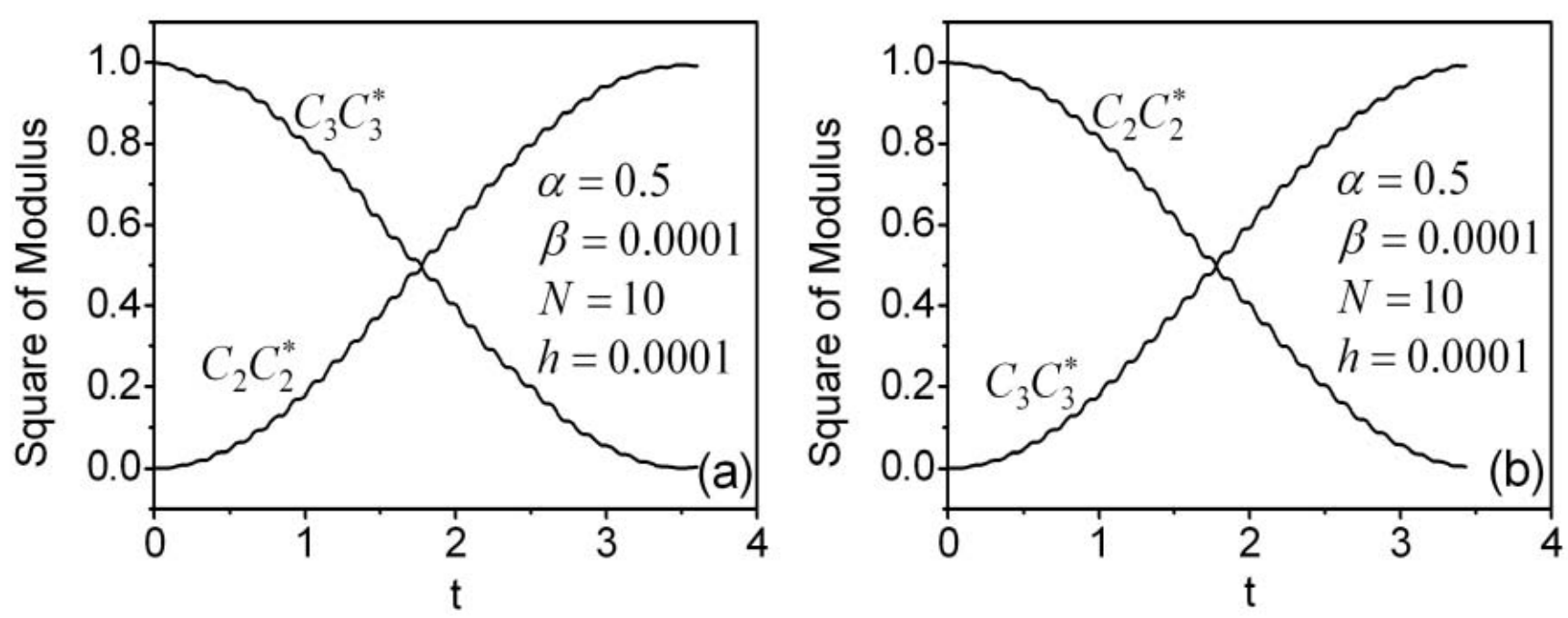

**Figure 5.** Stimulated transition between two stationary states $\phi_2(x)$ and $\phi_3(x)$, determined by the NLSE with a periodical potential $V(x,t) = \sum_{n=1}^{N} \phi_n(x)\cos[(E_3 - E_2)t]$, where $\phi_n(x) = \sqrt{2}\sin(n\pi x)$ and $E_n = 0.5(n\pi)^2$ with $n = 1,2,\cdots,N$ are respectively the eigenfunctions and eigenvalues of the linear 1D ISWP Hamiltonian. Graphs (a) and (b) respectively correspond to the transition from $\phi_3(x)$ to $\phi_2(x)$ and the transition from $\phi_2(x)$ to $\phi_3(x)$.

The NLSE expressed in Eq. (26) with a small positive $\beta$ may thus present both the stimulated transition and the spontaneous transition between stationary states. For the stimulated transition, the external potential dominates the wave-function evolution. For the spontaneous transition, however, the nonlinear term plays a key role because LSE without this term cannot describe this transition. The spontaneous transition is dominated jointly by the nonlinear term and the perturbation potential: Without the nonlinear term, the wave function will not necessarily evolve into a stationary state; without the perturbation potential, the wave function will remain in the initial stationary state. The relative magnitudes of the perturbation potential and the nonlinear term determine the final state of the transition. Both stimulated transition and spontaneous transition are the result of the competition between the external potential and the nonlinear term.

## 6. Conclusions

In conclusion, the inexactness of the standard LSE indicates the necessity of an NLSE that contains the time derivative of the probability density. One such NLSE is investigated because it has the simplest form and maintains important properties of the corresponding LSE. For the 1D ISWP and some time-dependent potentials that tend to the 1D ISWP, numerical calculations demonstrate that this simplest NLSE presents the state evolution similar to the wave-function reduction, because the wave function always evolves into an eigenstate of the linear Hamiltonian of the 1D ISWP. This NLSE may be an approximation of a more complicated and more exact equation of dynamics, and tentative calculations indicate that its realization of the wave-function reduction may be a universal conclusion valid for any potentials. The nonlinear term containing the time derivative of the probability density may provide clues to the solutions of some unsolved problems in quantum mechanics.

## References

[1] von Neumann J 1955 *Mathematical Foundations of Quantum Mechanics* (Princeton: Princeton University Press)
[2] Dirac P 1958 *The Principles of Quantum Mechanics* (Oxford: Clarendon Press)
[3] Zurek W H 2003 *Rev. Mod. Phys.* **75** 715
[4] Ghirardi G C, Rimini A and Weber T 1986 *Phys. Rev. D* **34** 470
[5] Ghirardi G C, Pearle P and Rimini A 1990 *Phys. Rev. A* **42** 78
[6] Adler S L and Bassi A 2009 *Science* **325** 275
[7] Pearle P 2005 *Phys. Rev. A* **72** 022112
[8] Pearle P 1976 *Phys. Rev. D* **13** 857
[9] Bächtold M 2008 *J. Gen. Philos. Sci.* **39** 17
[10] Shen Y R 1984 *The principles of nonlinear optics* (New York: Wiley)
[11] Gross E 1961 *Nuovo Cimento* **20** 454
[12] Gross E 1963 *J. Math. Phys.* **4** 195
[13] Pitaevskii L P 1961 *Zh. Eksp. Teor. Fiz.* **40** 646 (1961 *Sov. Phys. JETP* **13** 451)
[14] Salerno M, Konotop V V and Bludov Y V 2008 *Phys. Rev. Lett.* **101** 030405
[15] Arevalo E 2009 *Phys. Rev. Lett.* **102** 224102
[16] Buryak A V, Trapani P D, Skryabin D V and Trillo S 2002 *Phys. Rep.* **370** 63
[17] Eisenberg H S, Silberberg Y, Morandotti R, Boyd A R and Aitchison J S 1998 *Phys. Rev. Lett.* **81** 3383
[18] Lederer F, Stegeman G I, Christodoulides D N, Assanto G, Segev M and Silberberg Y 2008 *Phys. Rep.* **463** 1
[19] Shukla P K and Eliasson B 2007 *Phys. Rev. Lett.* **99** 096401
[20] Dalfovo F, Giorgini S, Pitaevskii L P and Stringari S 1999 *Rev. Modern Phys.* **71** 463
[21] Ruprecht P A, Holland M J, Burnett K and Edwards M 1995 *Phys. Rev. A* **51** 4704
[22] Adhikari S K 2000 *Phys. Rev. E* **62** 2937
[23] Parwani R and Tabia G 2007 *J. Phys. A: Math. Theor.* **40** 5621
[24] Lashkin V M 2007 *J. Phys. A: Math. Theor.* **40** 6119
[25] Florjanczyk M and Gagnon L 1990 *Phys. Rev. A* **41** 4478
[26] Bohm D 1951 *Quantum Theory* (New York: Prentice-Hall Incorporation)
[27] Jackson J D 1999 *Classical Electrodynamics* (New York: Wiley)